\documentclass[prc,twocolumn,aps,superscriptaddress,showpacs]{revtex4}
\usepackage{amssymb}
\usepackage{amsmath,bm}
\usepackage{graphicx}
\usepackage[normalem]{ulem}
\usepackage[dvips]{color}

\renewcommand\sout{\bgroup \color{red} \ULdepth=-.5ex \ULset}

\begin{document}

\title{Nuclear matter symmetry energy and the symmetry energy coefficient in
the mass formula}
\author{Lie-Wen Chen}
\affiliation{Department of Physics, Shanghai Jiao Tong University, Shanghai 200240, China}
\affiliation{Center of Theoretical Nuclear Physics, National Laboratory of Heavy Ion
Accelerator, Lanzhou 730000, China}
\date{\today}

\begin{abstract}
Within the Skyrme-Hartree-Fock (SHF) approach, we show that for a fixed mass
number $A$, both the symmetry energy coefficient $a_{\text{\textrm{sym}}}(A)$
in the semi-empirical mass formula and the nuclear matter symmetry energy $%
E_{\text{\textrm{sym}}}({\rho _{A}})$ at a subsaturation reference density $%
\rho _{A}$ can be determined essentially by the symmetry energy $E_{\text{%
\textrm{sym}}}({\rho _{0}})$ and its density slope $L$ at saturation density
${\rho _{0}}$. Meanwhile, we find the dependence of $a_{\text{\textrm{sym}}%
}(A)$ on $E_{\text{\textrm{sym}}}({\rho _{0}})$ or $L$ is approximately
linear and is very similar to the corresponding linear dependence displayed
by $E_{\text{\textrm{sym}}}({\rho _{A}})$, providing an explanation for the
relation $E_{\text{\textrm{sym}}}({\rho _{A}})\approx a_{\text{\textrm{sym}}%
}(A)$. Our results indicate that a value of
$E_{\text{\textrm{sym}}}({\rho _{A}})$ leads to a linear correlation
between $E_{\text{\textrm{sym}}}({\rho _{0}} )$ and $L$ and thus can
put important constraints on $E_{\text{\textrm{sym}}}({\rho _{0}} )$
and $L$. Particularly, the values of $E_{\text{\textrm{sym}}}(\rho
_{0})=$ $30.5\pm 3$ MeV and $L=$ $52.5\pm 20$ MeV are simultaneously
obtained by combining the constraints from recently extracted
$E_{\text{\textrm{sym}}}({\rho _{A}=0.1}$ {fm}$^{{-3}})$ with those
from recent analyses of neutron skin thickness of Sn isotopes in the
same SHF approach.
\end{abstract}

\pacs{21.65.Ef, 21.30.Fe, 21.10.Gv, 21.60.Jz}
\maketitle

\section{Introduction}

The study of the nuclear matter symmetry energy $E_{\text{\textrm{sym}}%
}(\rho )$, which essentially characterizes the isospin dependent part of the
equation of state (EOS) of asymmetric nuclear matter, is currently an
exciting topic of research in nuclear physics. Knowledge about the symmetry
energy is essential in understanding many aspects of nuclear physics and
astrophysics~\cite{LiBA98,Dan02,Lat04,Ste05,Bar05,LCK08} as well as some
interesting issues regarding possible new physics beyond the standard model
\cite{Hor01b,Sil05,Kra07,Wen09}. In recent years, significant progress has
been made in determining the density dependence of $E_{\text{\textrm{sym}}%
}(\rho )$ \cite{Bar05,LCK08}, especially its value $E_{\text{\textrm{sym}}}({%
\rho _{0}})$ and its density slope $L$ at saturation density ${\rho _{0}}$.
While constraints on $E_{\text{\textrm{sym}}}(\rho _{0})$ and $L$ from
different experimental data or methods become consistently convergent \cite%
{Tsa09,Car10,She10,XuC10}, they are still far from an accuracy required for
understanding enough precisely many important properties of neutron stars
\cite{Lat04,XuJ09}. To narrow the uncertainty of the constrains on $E_{\text{%
\textrm{sym}}}(\rho _{0})$ and $L$ by using more accurate data or new
methods is thus of crucial importance.

Recently, based on the calculations within the droplet model and mean field
models using a number of different parameter sets, Centelles \textsl{et al.}
found that the symmetry energy coefficient $a_{\text{\textrm{sym}}}(A)$ of
finite nuclei with mass number $A$ in the semi-empirical mass formula can
approximately equal to nuclear matter symmetry energy $E_{\text{\textrm{sym}}%
}({\rho _{A}})$ at a reference density $\rho _{A}$ in the subnormal density
region, i.e., $E_{\text{\textrm{sym}}}({\rho _{A}})\approx a_{\text{\textrm{%
sym}}}(A)$ \cite{Cen09}. This relation provides the possibility to directly
determine the symmetry energy at subnormal densities from the semi-empirical
mass formula and also has many important implications for extracting the
symmetry energy from isospin dependent observables of finite nuclei \cite%
{Cen09}. While this relation has been used to extract information on the
symmetry energy around the normal density \cite{Cen09,Liu10}, its
microscopic explanation is still missing.

Within the Skyrme-Hartree-Fock (SHF) energy density functional, we
demonstrate in the present work that both $a_{\text{\textrm{sym}}}(A)$ and $%
E_{\text{\textrm{sym}}}({\rho _{A}})$ can be determined essentially by $E_{%
\text{\textrm{sym}}}(\rho _{0})$ and $L$ and meanwhile they display very
similar linear dependence on $E_{\text{\textrm{sym}}}(\rho _{0})$ or $L$,
and thus providing an explanation for the relation $E_{\text{\textrm{sym}}}({%
\rho _{A}})\approx a_{\text{\textrm{sym}}}(A)$. Furthermore, we show that a
value of $E_{\text{\textrm{sym}}}({\rho _{A}})$ can put important
constraints on $E_{\text{\textrm{sym}}}({\rho _{0}})$ and $L$. Combining the
constraints from recently extracted $E_{\text{\textrm{sym}}}({\rho _{A}=0.1}$
{fm}$^{{-3}})$ with those from recent analyses of existing data on neutron
skin thickness of Sn isotopes \cite{Che10} within the same SHF approach
leads to stringent constraints simultaneously on $E_{\text{\textrm{sym}}%
}(\rho _{0})$ and $L$.

\section{$E_{\text{\textrm{sym}}}({\protect\rho })$ and $a_{\text{\textrm{sym%
}}}(A)$ in the Skyrme-Hartree-Fock approach}

\label{Theory}

The EOS of isospin asymmetric nuclear matter, given by its binding energy
per nucleon, can be expanded to $2$nd-order in isospin asymmetry $\delta $ as%
\begin{equation}
E(\rho ,\delta )=E_{0}(\rho )+E_{\mathrm{sym}}(\rho )\delta ^{2}+O(\delta
^{4}),  \label{EOSANM}
\end{equation}%
where $\rho =\rho _{n}+\rho _{p}$ is the baryon density with $\rho _{n}$ and
$\rho _{p}$ denoting the neutron and proton densities, respectively; $\delta
=(\rho _{n}-\rho _{p})/\rho $ is the isospin asymmetry; $E_{0}(\rho )=E(\rho
,\delta =0)$ is the binding energy per nucleon in symmetric nuclear matter,
and the nuclear symmetry energy is expressed as%
\begin{eqnarray}
E_{\mathrm{sym}}(\rho ) &=&\frac{1}{2!}\frac{\partial ^{2}E(\rho ,\delta )}{%
\partial \delta ^{2}}|_{\delta =0}  \notag \\
&=&E_{\mathrm{sym}}(\rho _{0})+L\chi +\frac{K_{\mathrm{sym}}}{2!}\chi
^{2}+O(\chi ^{3}),
\end{eqnarray}%
with $\chi =\frac{\rho -\rho _{0}}{3\rho _{0}}$. The coefficients $L=3\rho
_{0}\frac{dE_{\mathrm{sym}}(\rho )}{d\rho }|_{\rho =\rho _{0}}$ and $K_{%
\mathrm{sym}}=9\rho _{0}^{2}\frac{d^{2}E_{\mathrm{sym}}(\rho )}{d\rho ^{2}}%
|_{\rho =\rho _{0}}$ are the slope and curvature parameters of the symmetry
energy, respectively. Within the standard SHF approach, the symmetry energy
can be written as (see, e.g., Ref. \cite{Cha97})%
\begin{eqnarray}
E_{\text{\textrm{sym}}}(\rho ) &=&\frac{\hbar ^{2}}{6m}\left( \frac{3\pi ^{2}%
}{2}\right) ^{2/3}\rho ^{\frac{2}{3}}-\frac{1}{8}t_{0}(2x_{0}+1)\rho  \notag
\\
&-&\frac{1}{24}\left( \frac{3\pi ^{2}}{2}\right) ^{2/3}\Theta _{\mathrm{sym}%
}\rho ^{\frac{5}{3}}  \notag \\
&-&\frac{1}{48}t_{3}(2x_{3}+1)\rho ^{\sigma +1},
\end{eqnarray}%
with $\Theta _{\mathrm{sym}}=3t_{1}x_{1}-t_{2}(4+5x_{2})$ and $\sigma $, $%
t_{0}-t_{3}$, $x_{0}-x_{3}$ being the Skyrme interaction parameters.

As shown in Refs.~\cite{Che10,Che11}, the $9$ Skyrme interaction parameters,
i.e., $\sigma $, $t_{0}-t_{3}$, $x_{0}-x_{3}$ can be expressed analytically
in terms of $9$ macroscopic quantities $\rho _{0}$, $E_{0}(\rho _{0})$, the
incompressibility $K_{0}$, the isoscalar effective mass $m_{s,0}^{\ast }$,
the isovector effective mass $m_{v,0}^{\ast }$, $E_{\text{\textrm{sym}}}({%
\rho _{0}})$, $L$, gradient coefficient $G_{S}$, and symmetry-gradient
coefficient $G_{V}$. In terms of these macroscopic quantities, the symmetry
energy can be rewritten as

\begin{equation}
E_{\text{\textrm{sym}}}(\rho )=A_{1}E_{\text{\textrm{sym}}}({\rho _{0}}%
)+B_{1}L+C_{1},  \label{EsymLinear}
\end{equation}%
with%
\begin{eqnarray}
A_{1} &=&(\gamma u-u^{\gamma })/(\gamma -1) \\
B_{1} &=&(u^{\gamma }-u)/[3(\gamma -1)] \\
C_{1} &=&E_{\text{\textrm{sym}}}^{\mathrm{kin}}({\rho _{0}})u^{2/3}+Du^{5/3}
\notag \\
&&-\frac{(3\gamma -2)E_{\text{\textrm{\ sym}}}^{\mathrm{kin}}({\rho _{0}}%
)+(3\gamma -5)D}{3(\gamma -1)}u  \notag \\
&&+\frac{E_{\text{\textrm{\ sym}}}^{\mathrm{kin}}({\rho _{0}})-2D}{3(\gamma
-1)}u^{\gamma },
\end{eqnarray}%
where $u={\rho /}\rho _{0}$ is the reduced density; $E_{\text{\textrm{sym}}%
}^{kin}({\rho _{0}})=\frac{\hbar ^{2}}{6m}\left( \frac{3\pi ^{2}}{2}{\rho
_{0}}\right) ^{2/3}$ is the kinetic symmetry energy at ${\rho _{0}}$; and
the parameters $D$ and $\gamma $ are defined as~\cite{Che09a,Che09b}

\begin{eqnarray}
D &=&\frac{5}{9}E_{\mathrm{kin}}^{0}\left( 4\frac{m}{m_{s,0}^{\ast }}-3\frac{%
m}{m_{v,0}^{\ast }}-1\right) \\
\gamma &=&\sigma +1=\frac{K_{0}+2E_{\mathrm{kin}}^{0}-10C}{3E_{\mathrm{kin}%
}^{0}-9E_{0}(\rho _{0})-6C},
\end{eqnarray}%
with $C=\frac{m-m_{s,0}^{\ast }}{m_{s,0}^{\ast }}E_{\mathrm{kin}}^{0}$ and $%
E_{\mathrm{kin}}^{0}=\frac{3\hbar ^{2}}{10m}\left( \frac{3}{2}\pi ^{2}\rho
_{0}\right) ^{2/3}$.

The symmetry energy coefficient $a_{\text{\textrm{sym}}}(A)$ of finite
nuclei in the semi-empirical mass formula can be expressed as \cite{Cen09}

\begin{equation}
a_{\text{\textrm{sym}}}(A)=\frac{E_{\text{\textrm{sym}}}({\rho _{0}})}{%
1+x_{A}},\text{ with }x_{A}=\frac{9E_{\text{\textrm{sym}}}({\rho _{0}})}{4Q}%
A^{-1/3},  \label{asymA}
\end{equation}%
where the $Q$ parameter is the so-called neutron-skin stiffness coefficient
in the droplet model \cite{Mye69,Bra85} and it is related to the nuclear
surface symmetry energy \cite{Dan03,Dan09}. Usually for a given nuclear
interaction, the $Q$ parameter can be obtained from asymmetric semi-infinite
nuclear matter calculations \cite{Mye69,Bra85,Tre86,Cen98,Dan09}. As a good
approximation, the $Q$ parameter can be expressed as \cite{Tre86}
\begin{equation}
Q=\frac{9}{4}\frac{E_{\text{\textrm{sym}}}^{2}({\rho _{0}})}{\varepsilon
_{\delta }^{e}},\text{ with }\varepsilon _{\delta }^{e}=\frac{2a}{r_{\text{%
\textrm{nm}}}}\left( L-\frac{K_{\text{\textrm{sym}}}}{12}\right) ,
\end{equation}%
where $r_{\text{\textrm{nm}}}=\left( \frac{4}{3}\pi {\rho _{0}}\right)
^{-1/3}$ is the radius constant of nuclear matter and $a$ is the diffuseness
parameter in the Fermi-like function from the parametrization of nuclear
surface profile of symmetric semi-infinite nuclear matter. Many calculations
\cite{Tre86,Dan03,Dan09} have indicated $a\approx 0.55$ fm and then $%
2a/r_{\textrm{nm}}\approx 1$. Therefore, the $x_{A}$ parameter can
be approximated by
\begin{equation}
x_{A}=\left( L-K_{\text{\textrm{sym}}}/12\right) \frac{A^{-1/3}}{E_{\text{%
\textrm{sym}}}({\rho _{0}})}.  \label{xA}
\end{equation}%
It should be noted that Eq. (\ref{xA}) is a good approximation for
evaluating the $a_{\text{\textrm{sym}}}(A)$, and within the standard
SHF energy density functional the difference between the value of
$a_{\text{\textrm{sym}}}(A=208)$ from Eq.(\ref{xA}) and that of
using the exact $Q$ parameter obtained from asymmetric semi-infinite
nuclear matter calculations is essentially less than $1$ MeV
\cite{Che11b}. Furthermore, within the standard SHF approach,
$K_{\text{\textrm{sym}}}$ can be written
in terms of the macroscopic quantites as \cite{Che09b}%
\begin{eqnarray}
K_{\text{\textrm{sym}}} &=&3\gamma L+E_{\text{\textrm{sym}}}^{kin}({\rho _{0}%
})(3\gamma -2)  \notag \\
&&+2D(5-3\gamma )-9\gamma E_{\text{\textrm{sym}}}({\rho _{0}}),
\label{KsymLSHF}
\end{eqnarray}%
and thus we have

\begin{eqnarray}
x_{A} &=&[\frac{4-\gamma }{4}L+\frac{3}{4}\gamma E_{\text{\textrm{sym}}}({%
\rho _{0}})-\frac{(3\gamma -2)}{12}E_{\text{\textrm{sym}}}^{kin}({\rho _{0}})
\notag \\
&&-\frac{(5-3\gamma )}{6}D]\frac{A^{-1/3}}{E_{\text{\textrm{sym}}}({\rho _{0}%
})}.  \label{xASHF}
\end{eqnarray}%
For $|x_{A}|<1$, $a_{\text{\textrm{sym}}}(A)$ in Eq. (\ref{asymA}) can be
expanded as%
\begin{equation}
a_{\text{\textrm{sym}}}(A)=E_{\text{\textrm{sym}}}({\rho _{0}}%
)(1-x_{A}+x_{A}^{2}-\cdot \cdot \cdot )\text{.}  \label{asymAExpand}
\end{equation}%
Neglectiing the $x_{A}^{2}$ and higher-order terms in Eq. (\ref{asymAExpand}%
) leads to

\begin{equation}
a_{\text{\textrm{sym}}}(A)=A_{2}E_{\text{\textrm{sym}}}({\rho _{0}}%
)+B_{2}L+C_{2},  \label{asymALinear1}
\end{equation}%
with%
\begin{eqnarray}
A_{2} &=&\left( 1-\frac{3\gamma }{4}A^{-1/3}\right)  \\
B_{2} &=&-\frac{4-\gamma }{4}A^{-1/3} \\
C_{2} &=&\left( \frac{(3\gamma -2)}{12}E_{\text{\textrm{sym}}}^{kin}({\rho
_{0}})+\frac{(5-3\gamma )}{6}D\right) A^{-1/3}.
\end{eqnarray}%
However, the convergence of the expansion in Eq. (\ref{asymAExpand}) is
usually very slow and thus Eq. (\ref{asymALinear1}) is a very bad
approximation to\ $a_{\text{\textrm{sym}}}(A)$ even for heavy nuclei \cite%
{Dan03,Rei06}. A much better approximation could be obtained by the
two-variable Taylor expansion with respect to $E_{\text{\textrm{sym}}}({\rho
_{0}})$ and $L$ at a point of $E_{\text{\textrm{sym}}}({\rho _{0}})=S_{0}$
and $L=L_{0}$ as%
\begin{eqnarray}
a_{\text{\textrm{sym}}}(A) &=&a_{\text{\textrm{sym}}}(A)|_{E_{\text{\textrm{%
sym}}}({\rho _{0}})=S_{0},L=L_{0}}  \notag \\
&+&\left( E_{\text{\textrm{sym}}}({\rho _{0}})-S_{0}\right) \frac{\partial
a_{\text{\textrm{sym}}}(A)}{\partial E_{\text{\textrm{sym}}}({\rho _{0}})}%
|_{E_{\text{\textrm{sym}}}({\rho _{0}})=S_{0},L=L_{0}}  \notag \\
&+&\left( L-L_{0}\right) \frac{\partial a_{\text{\textrm{sym}}}(A)}{\partial
L}|_{E_{\text{\textrm{sym}}}({\rho _{0}})=S_{0},L=L_{0}}+\cdot \cdot \cdot
\end{eqnarray}%
and keeping only the first-order terms leads to%
\begin{equation}
a_{\text{\textrm{sym}}}(A)=A_{3}E_{\text{\textrm{sym}}}({\rho _{0}}%
)+B_{3}L+C_{3},  \label{asymALinear2}
\end{equation}%
with%
\begin{eqnarray}
A_{3} &=&\frac{1-\frac{3}{4}\gamma A^{-1/3}+2x_{A}^{0}}{(1+x_{A}^{0})^{2}} \\
B_{3} &=&\frac{\frac{\gamma -4}{4}A^{-1/3}}{(1+x_{A}^{0})^{2}} \\
C_{3} &=&\frac{S{_{0}}}{1+x_{A}^{0}}-S_{0}A_{3}-L_{0}B_{3}
\end{eqnarray}%
and%
\begin{eqnarray}
x_{A}^{0} &=&[\frac{3}{4}\gamma S{_{0}}+\frac{4-\gamma }{4}L_{0}-\frac{%
(3\gamma -2)}{12}E_{\text{\textrm{sym}}}^{kin}({\rho _{0}})  \notag \\
&&-\frac{(5-3\gamma )}{6}D]\frac{A^{-1/3}}{S{_{0}}}.
\end{eqnarray}

\section{Numerical results and discussions}

\label{Result}

One can see from Eq. (\ref{EsymLinear}) that $E_{\text{\textrm{sym}}}({\rho }%
)$ is linear functions of $E_{\text{\textrm{sym}}}({\rho _{0}})$ and $L$
with the coefficients $A_{1}$, $B_{1}$ and $C_{1}$ determined by the density
$\rho $ and nuclear matter macroscopic quantities $\rho _{0}$, $E_{0}(\rho
_{0})$, $K_{0}$, $m_{s,0}^{\ast }$, and $m_{v,0}^{\ast }$. Meanwhile, from
Eqs. (\ref{asymA}) and (\ref{xASHF}), one can see that $a_{\text{\textrm{sym}%
}}(A)$ is determined by the mass number $A$ and also the nuclear matter
macroscopic quantities. In particular, $a_{\text{\textrm{sym}}}(A)$ can also
be linear functions of $E_{\text{\textrm{sym}}}({\rho _{0}})$ and $L$ if the
approximation (\ref{asymALinear1}) or (\ref{asymALinear2}) is valid. As
mentioned previously, the relation $a_{\text{\textrm{sym}}}(A)\approx E_{%
\text{\textrm{sym}}}({\rho _{A}})$ has been observed within mean field
models using a number of different parameter sets for the nuclear effective
interactions. Particularly, one finds $a_{\text{\textrm{sym}}}(A=208)\approx
E_{\text{\textrm{sym}}}({\rho _{A}}=0.1$ fm$^{-3})$, $a_{\text{\textrm{sym}}%
}(A=116)\approx E_{\text{\textrm{sym}}}({\rho _{A}}=0.093$ fm$^{-3})$, and $%
a_{\text{\textrm{sym}}}(A=40)\approx E_{\text{\textrm{sym}}}({\rho _{A}}%
=0.08 $ fm$^{-3})$ \cite{Cen09}. This feature implies that $a_{\text{\textrm{%
sym}}}(A)$ and $E_{\text{\textrm{sym}}}({\rho _{A}})$ would display similar
correlation with each nuclear matter macroscopic quantity among $L$, $G_{V}$%
, $G_{S}$, $E_{0}(\rho _{0})$, $E_{\text{\textrm{sym}}}(\rho _{0})$, $K_{0}$%
, $m_{s,0}^{\ast }$, $m_{v,0}^{\ast }$, and $\rho _{0}$, which completely
determine the $9$ Skyrme interaction parameters $\sigma $, $t_{0}-t_{3}$, $%
x_{0}-x_{3}$. In the following, we show that this is indeed the case by
analyzing the correlations of $E_{\text{\textrm{sym}}}(\rho =0.1$\ fm$^{-3})$%
\ and $a_{\text{\textrm{sym}}}(A=208)$ with the nuclear matter macroscopic
quantities. We have also checked the cases of $A=116$ and $40$, and obtained
the similar conclusion as in the case of $A=208$ and confirmed the relations
$a_{\text{\textrm{sym}}}(A=116)\approx E_{\text{\textrm{sym}}}({\rho _{A}}%
=0.093$ fm$^{-3})$ and $a_{\text{\textrm{sym}}}(A=40)\approx E_{\text{%
\textrm{sym}}}({\rho _{A}}=0.08$ fm$^{-3})$.

As a reference for the correlation analyses based on the standard SHF energy
density functional, we use in the present work the MSL0 parameter set~\cite%
{Che10}, which is obtained by using the following empirical values for the
macroscopic quantities: $\rho _{0}=0.16$ fm$^{-3}$, $E_{0}(\rho _{0})=-16$
MeV, $K_{0}=230$ MeV, $m_{s,0}^{\ast }=0.8m$, $m_{v,0}^{\ast }=0.7m$, $E_{%
\text{\textrm{sym}}}({\rho _{0}})=30$ MeV, and $L=60$ MeV, $G_{V}=5$ MeV$%
\cdot $fm$^{5}$, and $G_{S}=132$ MeV$\cdot $fm$^{5}$. And the spin-orbit
coupling constant $W_{0}=133.3$ MeV $\cdot $fm$^{5}$ is used to fit the
neutron $p_{1/2}-p_{3/2}$ splitting in $^{16}$O. It has been shown~\cite%
{Che10} that the MSL0 interaction can describe reasonably the binding
energies and charge rms radii for a number of closed-shell or
semi-closed-shell nuclei. It should be pointed out that the MSL0 is only
used here as a reference for the correlation analyses. Using other Skyrme
interactions obtained from fitting measured binding energies and charge rms
radii of finite nuclei as in usual Skyrme parametrization will not change
our conclusion.
\begin{figure}[tbp]
\includegraphics[scale=0.75]{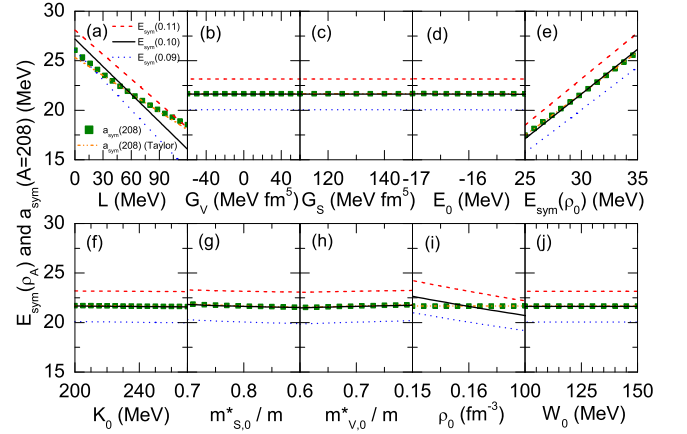}
\caption{{\protect\small (Color online) }$E_{\text{\textrm{sym}}}(\protect%
\rho _{A})${\protect\small \ with }$\protect\rho _{A}=0.09,0.10,$%
{\protect\small \ and }$0.11${\protect\small \ fm}$^{-3}${\protect\small \
as well as }$a_{\text{\textrm{sym}}}(A=208)${\protect\small \ and its
approximation with Taylor expansion in Eq. (\protect\ref{asymALinear2})
(with }$S_{0}=30${\protect\small \ MeV and }$L_{0}=60${\protect\small \
MeV)\ from SHF with MSL0 by varying individually }$L${\protect\small \ (a), }%
$G_{V}${\protect\small \ (b), }$G_{S}${\protect\small \ (c), }$E_{0}(\protect%
\rho _{0})${\protect\small \ (d), }$E_{\text{\textrm{sym}}}(\protect\rho %
_{0})${\protect\small \ (e), }$K_{0}${\protect\small \ (f), }$m_{s,0}^{\ast
} ${\protect\small \ (g), }$m_{v,0}^{\ast }${\protect\small \ (h), }$\protect%
\rho _{0}${\protect\small \ (i), and }$W_{0}${\protect\small \ (j).}}
\label{Xasym208}
\end{figure}

Shown in Fig. \ref{Xasym208} are $E_{\text{\textrm{sym}}}(\rho _{A})$ with $%
\rho _{A}=0.09,0.10,$ and $0.11$\ fm$^{-3}$\ as well as $a_{\text{\textrm{sym%
}}}(A=208)$\ (from Eq. (\ref{asymA}) with approximation in Eq.
(\ref{xA})) and its approximation with Taylor expansion in Eq.
(\ref{asymALinear2}) (with $S_{0}=30$ MeV and $L_{0}=60$ MeV)\
obtained from SHF with MSL0 by varying individually $L$, $G_{V}$,
$G_{S}$, $E_{0}(\rho _{0})$,
$E_{\text{\textrm{sym}}}(\rho _{0})$, $K_{0}$, $m_{s,0}^{\ast }$, $%
m_{v,0}^{\ast }$, $\rho _{0}$, and $W_{0}$, namely, varying one
quantity at a time while keeping all others at their default values
in MSL0. We note here that using the exact $Q$ parameter obtained
from asymmetric semi-infinite nuclear matter calculations to
evaluate the $a_{\text{\textrm{sym}}}(A)$ does not change our
conclusions \cite{Che11b}. It is interesting to see that, within the
uncertain ranges considered here for the macroscopic quantities,
$a_{\text{\textrm{sym}}}(A=208)$ displays strong correlations with
both $E_{\text{\textrm{sym}}}({\rho _{0}})$ and $L$ while
it is almost no dependence on other macroscopic quantities (From Eqs. (\ref%
{asymA}) and (\ref{xASHF}), $a_{\text{\textrm{sym}}}(A)$ is independent of $%
G_{V}$, $G_{S}$, and $W_{0}$). This is understandable since $a_{\text{%
\textrm{sym}}}(A)$ is determined uniquely by the three lowest-order
characteristic parameters of the symmetry energy, i.e., $E_{\text{\textrm{sym%
}}}(\rho _{0})$, $L$, and $K_{\text{\textrm{sym}}}$ as seen in Eqs. (\ref%
{asymA}) and (\ref{xA}) while $K_{\text{\textrm{sym}}}$ has been found to
strongly correlate with $E_{\text{\textrm{sym}}}(\rho _{0})$ and $L$ but
exhibit very weak dependence on other macroscopic quantities within the
standard SHF energy density functional as shown in Ref.\ \cite{Che11}.
Furthermore, $a_{\text{\textrm{sym}}}(A=208)$ displays approximately linear
correlations with both $E_{\text{\textrm{sym}}}(\rho _{0})$ and $L$ which is
demonstrated by the good approximation of Eq. (\ref{asymALinear2}) to $a_{%
\text{\textrm{sym}}}(A=208)$ observed in Fig. \ref{Xasym208}.

Similarly, one can see from Fig. \ref{Xasym208} that $E_{\text{\textrm{sym}}%
}(\rho _{A})$ with $\rho _{A}=0.09,0.10,$ and $0.11$\ fm$^{-3}$ display
strong linear correlations with both $E_{\text{\textrm{sym}}}({\rho _{0}})$
and $L$ while they are almost independent of other macroscopic quantities
except with a small dependence on $\rho _{0}$ (Note $E_{\text{\textrm{sym}}%
}(\rho _{A})$ is independent of $G_{V}$, $G_{S}$, and $W_{0}$). These
results indicate that $a_{\text{\textrm{sym}}}(A=208)$ and $E_{\text{\textrm{%
sym}}}({\rho _{A}})$ at subsaturation reference densities $\rho
_{A}=0.09,0.10,$ and $0.11$\ fm$^{-3}$ are essentially determined by $E_{%
\text{\textrm{sym}}}({\rho _{0}})$ and $L$. Furthermore, Fig. \ref{Xasym208}
shows that both $a_{\text{\textrm{sym}}}(A=208)$ and $E_{\text{\textrm{sym}}%
}({\rho _{A}})$ display very similar linear dependence on $E_{\text{\textrm{%
sym}}}({\rho _{0}})$ or $L$. Especially, it is seen from Fig. \ref{Xasym208}
that $E_{\text{\textrm{sym}}}(\rho _{A})$ with $\rho _{A}=0.10$\ fm$^{-3}$
gives the best fit to $a_{\text{\textrm{sym}}}(A=208)$. In other words, for
any Skyrme force determined by the $9$ parameters $\sigma $, $t_{0}-t_{3}$, $%
x_{0}-x_{3}$ or equivalently the $9$ macroscopic quantities $L$, $G_{V}$, $%
G_{S}$, $E_{0}(\rho _{0})$, $E_{\text{\textrm{sym}}}(\rho _{0})$, $K_{0}$, $%
m_{s,0}^{\ast }$, $m_{v,0}^{\ast }$, $\rho _{0}$, the value of $E_{\text{%
\textrm{sym}}}({\rho _{A}}=0.10\ $fm$^{-3})$ will approximately
equal to that of $a_{\text{\textrm{sym}}}(A=208)$, and this explains
the relation $E_{\text{\textrm{sym}}}({\rho _{A}}=0.10\
$fm$^{-3})\approx a_{\text{\textrm{sym}}}(A=208)$ observed within
mean field models using a number of different parameter sets.
Furthermore, one can see that the possible small deviations between
$E_{\text{\textrm{sym}}}({\rho _{A}}=0.10\ $fm$^{-3})$ and
$a_{\text{\textrm{sym}}}(A=208)$ observed for some different
parameter sets \cite{Cen09} may be mainly due to the different
values of $L$, $E_{\text{\textrm{sym}}}({\rho _{0}})$, and/or $\rho
_{0}$.
\begin{figure}[tbp]
\includegraphics[scale=0.85]{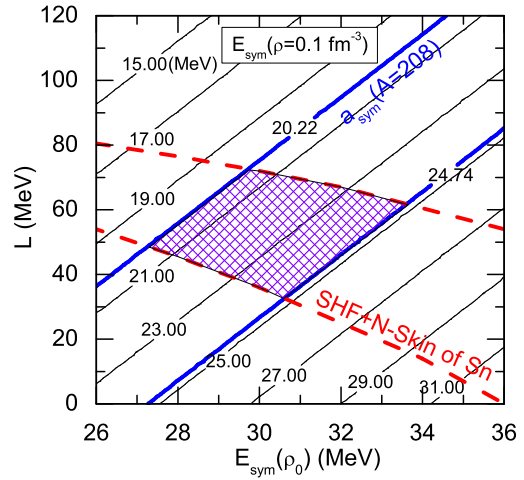}
\caption{{\protect\small (Color online) Contour curves in the }$E_{\text{%
\textrm{sym}}}(\protect\rho _{0})${\protect\small -}$L${\protect\small \
plane for }$E_{\text{\textrm{sym}}}(\protect\rho =0.1${\protect\small \ fm}$%
^{-3})${\protect\small . The region between the two thick solid lines
represents the constraint obtained in the present work with 20.22 MeV }$\leq
E_{\text{\textrm{sym}}}(\protect\rho _{A}=0.10\ ${\protect\small fm}$^{-3})$%
{\protect\small \ }$\leq ${\protect\small \ 24.74 MeV while the region
between the two thick dashed lines is the constraint from the SHF analysis
of neutron skin data of Sn isotopes within a }$2\protect\sigma $%
{\protect\small \ uncertainty \protect\cite{Che10}. The shaded region
represents the overlap of the two constraints.}}
\label{Esym01EsymL}
\end{figure}

Since $E_{\text{\textrm{sym}}}({\rho _{A}}=0.10\ $fm$^{-3})$ is essentially
determined by $E_{\text{\textrm{sym}}}({\rho _{0}})$ and $L$ and displays
linear correlations with the latters, a determination of $E_{\text{\textrm{%
sym}}}({\rho _{A}}=0.10\ $fm$^{-3})$ will then put important constraints on $%
E_{\text{\textrm{sym}}}({\rho _{0}})$ and $L$. As a matter of fact, the
value of the symmetry energy around $0.1$ fm$^{-3}$ has been heavily under
investigation in recent years in the literature \cite%
{Bro00,Hor01,Fur02,Pie02,Tod03,Vre03,Bal04,Col04,Yos04,Che05b,Sag07,Tri08,Cao08,Agr10,Jia10,Fat10}%
. For example, an analysis of the giant dipole resonance (GDR) of $^{208}$Pb
with Skyrme forces suggests a constraint $E_{\text{\textrm{sym}}}({\rho _{A}}%
=0.10\ $fm$^{-3})=23.3-24.9$ MeV \cite{Tri08} while a relativistic
mean-filed model analysis of the GDR of $^{132}$Sn leads to $E_{\text{%
\textrm{sym}}}({\rho _{A}}=0.10\ $fm$^{-3})=21.2-22.5$ MeV \cite{Cao08}. In
a recent work \cite{Liu10}, Liu \textsl{et al.} extracted the symmetry
energy coefficients $a_{\text{\textrm{sym}}}(A)$ for nuclei with mass number
$A=20-250$ from more than $2000$ measured nuclear masses and they obtained a
value of $20.22-24.74$ MeV for $a_{\text{\textrm{sym}}}(A=208)$ within a $%
2\sigma $ uncertainty and thus we have $E_{\text{\textrm{sym}}}({\rho _{A}}%
=0.10\ $fm$^{-3})=20.22-24.74$ MeV according to $E_{\text{\textrm{sym}}}({%
\rho _{A}}=0.10\ $fm$^{-3})\approx a_{\text{\textrm{sym}}}(A=208)$. In the
following, as a conservative estimate, we will use $E_{\text{\textrm{sym}}}({%
\rho _{A}}=0.10\ $fm$^{-3})=20.22-24.74$ MeV since it is essentially
consistent with the other two constraints from the GDR of $^{208}$Pb and $%
^{132}$Sn.

Shown in Fig. \ref{Esym01EsymL} are the contour curves in the $E_{\text{%
\textrm{sym}}}(\rho _{0})$-$L$\ plane for $E_{\text{\textrm{sym}}}(\rho =0.1$%
\ fm$^{-3})$ obtained from SHF with MSL0 by varying individually $E_{\text{%
\textrm{sym}}}(\rho _{0})$ and $L$. The region between the two thick solid
lines in Fig. \ref{Esym01EsymL} represents the constraint of $20.22$ MeV $%
\leq E_{\text{\textrm{sym}}}({\rho }=0.10\ $fm$^{-3})$\ $\leq $ $24.74$ MeV
from the nuclear mass restrictions on $a_{\text{\textrm{sym}}}(A=208)$. Also
included in Fig. \ref{Esym01EsymL} is a recent constraint (the region
between the two thick dashed lines) from SHF analysis of neutron skin data
of Sn isotopes within a $2\sigma $ uncertainty \cite{Che10}. It is
interesting to note that the constraint from the neutron skin data of Sn
isotopes suggests $L$ decreases with increasing $E_{\text{\textrm{sym}}%
}(\rho _{0})$ while the constraint from $E_{\text{\textrm{sym}}}({\rho }%
=0.10\ $fm$^{-3})$ displays opposite behaviors. This feature allows us to
extract simultaneously the values of both $E_{\text{\textrm{sym}}}(\rho
_{0}) $ and $L$ with higher accuracy, namely, $E_{\text{\textrm{sym}}}(\rho
_{0})=$ $30.5\pm 3$ MeV and $L=$ $52.5\pm 20$ MeV by combining the two
constraints as illustrated by the overlap of the two constraints represented
by the shaded region in Fig. \ref{Esym01EsymL}. Furthermore, it is seen from
Fig. \ref{Esym01EsymL} that a value of $E_{\text{\textrm{sym}}}({\rho }%
=0.10\ $fm$^{-3})$ puts a strong linear correlation between $E_{\text{%
\textrm{sym}}}(\rho _{0})$ and $L$, i.e., $L$ increases linearly with $E_{%
\text{\textrm{sym}}}(\rho _{0})$, which has been extensively observed in the
parameterization for nuclear effective interactions in mean-field models
\cite{Hor01,Fur02,Pie02,Tod03,Vre03,Bal04,Col04,Che05b,Agr10,Jia10,Fat10}.

The simultaneously extracted values of $E_{\text{\textrm{sym}}}(\rho _{0})=$
$30.5\pm 3$ MeV and $L=$ $52.5\pm 20$ MeV from the same SHF approach within
a $2\sigma $ uncertainty are essentially overlapped with other constraints
extracted from different experimental data or methods in the literature \cite%
{Mye69,Che05a,She07,Kli07,Tri08,Tsa09,Dan09,Cen09,Car10,XuC10,Liu10} (see
Ref. \cite{XuC10} for a recent summary) but with higher precision. In
particular, these extracted values are in remarkably good agreement with the
$E_{\text{\textrm{sym}}}(\rho _{0})=$ $31.3\pm 4.5$ MeV and $L=$ $52.7\pm
22.5$ MeV extracted most recently from global nucleon optical potentials
constrained by world data on nucleon-nucleus and (p,n) charge-exchange
reactions \cite{XuC10}. The extracted value of $L=$ $52.5\pm 20$ MeV also
agrees well with the value of $L=$ $58\pm 18$ MeV obtained from combining
the constraint from the neutron skin data of Sn isotopes \cite{Che10} with
those from recent analyses of isospin diffusion and the double
neutron/proton ratio in heavy-ion collisions at intermediate energies \cite%
{Tsa09}. Furthermore, the extracted value of $L=$ $52.5\pm 20$ MeV is
consistent with the value of $L=66.5$ MeV obtained from a recent systematic
analysis of the density dependence of nuclear symmetry energy within the
microscopic Brueckner-Hartree-Fock approach using the realistic Argonne V18
nucleon-nucleon potential plus a phenomenological three-body force of Urbana
type \cite{Vid09}. It should be stressed that the simultaneously extracted
values of $E_{\text{\textrm{sym}}}(\rho _{0})$ and $L$ in the present work
are obtained from the same Skyrme-Hartree-Fock energy density functional.

\section{Summary}

\label{Summary}

We have analyzed the correlations of the nuclear matter symmetry energy $E_{%
\text{\textrm{sym}}}({\rho })$ at a subsaturation reference density $\rho
_{A}$ and the symmetry energy coefficient $a_{\text{\textrm{sym}}}(A)$ of
finite nuclei in the semi-empirical mass formula with nuclear matter
macroscopic quantities\ within the Skyrme-Hartree-Fock energy density
functional. We have shown that $E_{\text{\textrm{sym}}}({\rho _{A}})$
displays explicitly linear correlations with $E_{\text{\textrm{sym}}}({\rho
_{0}})$ and $L$ and it is essentially determined by the latters but almost
no dependence on other macroscopic quantities except a small dependence on
the saturation density ${\rho _{0}}$. These features imply that a fixed
value of $E_{\text{\textrm{sym}}}({\rho _{A}})$ will lead to strong linear
correlation between $E_{\text{\textrm{sym}}}({\rho _{0}})$ and $L$.
Furthermore, we have found that the two macroscopic quantities $E_{\text{%
\textrm{sym}}}(\rho _{0})$ and $L$ essentially determine the value of $a_{%
\text{\textrm{sym}}}(A)$ and the latter displays approximately linear
correlations with both $E_{\text{\textrm{sym}}}({\rho _{0}})$ and $L$. In
particular, the correlation between $a_{\text{\textrm{sym}}}(A)$ and $E_{%
\text{\textrm{sym}}}({\rho _{0}})$ ($L$) is found to be very similar to that
between $E_{\text{\textrm{sym}}}({\rho _{A}})$ and $E_{\text{\textrm{sym}}}({%
\rho _{0}})$ ($L$), thus providing an explanation for the relation $E_{\text{%
\textrm{sym}}}({\rho _{A}})\approx a_{\text{\textrm{sym}}}(A)$ observed in
mean field models using a number of different parameter sets.

Using the relation $E_{\text{\textrm{sym}}}({\rho _{A}})\approx a_{\text{%
\textrm{sym}}}(A)$, we have demonstrated that within the Skyrme-Hartree-Fock
energy density functional, the value of $E_{\text{\textrm{sym}}}({\rho }%
=0.10\ $fm$^{-3})=20.22-24.74$ MeV extracted recently from nuclear masses
within a $2\sigma $ uncertainty \cite{Liu10} can put important constraints
on $E_{\text{\textrm{sym}}}({\rho _{0}})$ and $L$. Combining these
constraints with those from recent analyses of existing data on neutron skin
thickness of Sn isotopes\ based on the same Skyrme-Hartree-Fock approach
within a $2\sigma $ uncertainty \cite{Che10} allows us to extract
simultaneously the values of both $E_{\text{\textrm{sym}}}(\rho _{0})$ and $%
L $, i.e., $E_{\text{\textrm{sym}}}(\rho _{0})=$ $30.5\pm 3$ MeV and $L=$ $%
52.5\pm 20$ MeV. These extracted values are essentially consistent with
other constraints extracted from different experimental data in the
literature but with higher precision.

In the present work, all analyses are based on the standard
Skyrme-Hartree-Fock energy density functional. It will be interesting to see
how our results change if different energy-density functionals are used. On
the other hand, it will be also interesting to see how our results,
especially the new constraints on $E_{\text{\textrm{sym}}}(\rho _{0})$ and $%
L $ obtained in the present work, can give implications for the neutron-skin
thickness of heavy nuclei, the isovector giant dipole resonance of finite
nuclei, and properties of neutron stars. These studies are in progress.

\section*{ACKNOWLEDGMENTS}

This work was supported in part by the NNSF of China under Grant No.
10975097, Shanghai Rising-Star Program under grant No. 11QH1401100,
and the National Basic Research Program of China (973 Program) under
Contract No. 2007CB815004.

\end{document}